\documentclass[aps,reprint,prx,superscriptaddress]{revtex4-2}
\usepackage{algpseudocode,amsfonts,amsmath,amssymb,bm,dsfont,mathtools}

\usepackage{lipsum,xcolor}

\newcommand{\grad}{\nabla}

\DeclarePairedDelimiter{\norm}{||}{||}
\DeclareMathOperator*{\argmin}{argmin}
\DeclareMathOperator*{\Corr}{Corr}

\def\ba{\mathbf{a}}

\def\bc{\mathbf{c}}

\def\bff{\mathbf{f}}

\def\bk{\mathbf{k}}

\def\bn{\mathbf{n}}

\def\bp{\mathbf{p}}
\def\bq{\mathbf{q}}
\def\br{\mathbf{r}}

\def\bu{\mathbf{u}}

\def\bw{\mathbf{w}}
\def\bx{\mathbf{x}}
\def\by{\mathbf{y}}
\def\bz{\mathbf{z}}

\def\bB{\mathbf{B}}

\def\bD{\mathbf{D}}
\def\bE{\mathbf{E}}
\def\bF{\mathbf{F}}
\def\bG{\mathbf{G}}

\def\bI{\mathbf{I}}

\def\bQ{\mathbf{Q}}

\def\bS{\mathbf{S}}
\def\bT{\mathbf{T}}

\def\btheta{\bm{\theta}}

\def\blambda{\bm{\lambda}}

\def\bsigma{\bm{\sigma}}

\def\bPhi{\bm{\Phi}}

\def\bOmega{\bm{\Omega}}

\def\cF{\mathcal{F}}

\def\cH{\mathcal{H}}

\def\veps{\varepsilon}

\renewcommand{\div}{\grad\cdot}

\def\obu{\overline{\mathbf{u}}}

\def\obE{\overline{\mathbf{E}}}
\def\obQ{\overline{\mathbf{Q}}}
\def\obS{\overline{\mathbf{S}}}

\def\obSigma{\overline{\bm{\Sigma}}}

\def\trace{\rm trace}

\begin{document}

\title{Correlations, mean-field limits, and transition to the concentrated regime in motile particle suspensions}

\author{Bryce Palmer}
\affiliation{Department of Mechanical Engineering, Michigan State University, East Lansing, Michigan 48824, USA}

\author{Scott Weady}
\email{sweady@flatironinstitute.org}
\affiliation{Center for Computational Biology, Flatiron Institute, New York, New York 10010, USA}

\author{Michael O'Brien}
\affiliation{Center for Computational Biology, Flatiron Institute, New York, New York 10010, USA}
\affiliation{Department of Physics, Harvard University, Cambridge, Massachusetts 02138, USA}

\author{Blakesley Burkhart}
\affiliation{Center for Computational Astrophysics, Flatiron Institute, New York, New York 10010, USA}
\affiliation{Department of Physics and Astronomy, Rutgers University, Piscataway, New Jersey 08854, USA}

\author{Michael J. Shelley}
\email{mshelley@flatironinstitute.org}
\affiliation{Center for Computational Biology, Flatiron Institute, New York, New York 10010, USA}
\affiliation{Courant Institute of Mathematical Sciences, New York University, New York, New York 10012, USA}

\begin{abstract}
    Suspensions of swimming particles exhibit complex collective behaviors driven by hydrodynamic interactions, showing persistent large-scale flows and long-range correlations. While heavily studied, it remains unclear how such structures depend on the system size and swimmer concentration. To address these issues, we simulate very large systems of suspended swimmers across a range of system sizes and volume fractions. For this we use high-performance simulation tools that build on slender body theory and implicit resolution of steric interactions. At low volume fractions and long times, the particle simulations reveal dynamic flow structures and correlation functions that scale with the system size. These results are consistent with a mean-field limit and agree well with a corresponding kinetic theory. At higher concentrations, the system departs from mean-field behavior. Flow structures become cellular, and correlation lengths scale with the particle size. Here, translational motion is suppressed, while rotational dynamics dominate. These findings highlight the limitations of dilute mean-field models and reveal new behaviors in dense active suspensions.
\end{abstract}

\maketitle

\section{Introduction}

Suspensions of active particles, such as collections of motile bacteria \cite{dombrowski2004self,sokolov2007concentration} or microtubules mixed with molecular motors \cite{sanchez2012spontaneous,duclos2020topological}, are a canonical class of active matter. Driven at the particle scale, these systems can spontaneously organize into collective flows displaying persistent and fluctuating large-scale structures and long-range correlations of microstructural configurations \cite{gachelin2014collective,liu2021density,Peng2021}. It is particularly evident for suspensions of microswimmers that such collective dynamics is mediated predominately by hydrodynamic interactions in which particle motion and corresponding stresses are coupled through the surrounding fluid. A central question in these systems is how the microscopic details, such as  swimmer shape and propulsive mechanism, lead to the observed macroscopic behaviors. 

While systems composed of living components are highly complex, computational many-particle models can permit careful control of the microscopic physics albeit in much simplified form. In the most explicit case for microswimmers in a Newtonian fluid, such models take the form of a boundary value problem for the fluid velocity in which a body deformation, or yet more simply, a stress or slip velocity, is prescribed along the surface of each particle subject to force- and torque-free constraints. Directly solving such boundary value problems for many bodies remains exceptionally difficult, though advances have been made in this direction \cite{rahimian2010petascale,kohl2023fast,Malhotra2024}. As a consequence, particle models typically approximate particle-induced velocities by, for example, their far-field flow contributions \cite{Hernandez2005,Stenhammar2017,Bardfalvy2019}, perhaps coupled to near-field effects in denser suspensions \cite{mehandia2008collective,ishikawa2008development,evans2011orientational,elfring2022active}, or use simplified asymptotic geometries \cite{SS2007,SS2012}. Such models have been successful in describing many experimental observations, such as a volume fraction-dependent transition to instability and large-scale spatial correlations \cite{liu2021density,Peng2021}.

While many-particle simulations are the gold standard for theoretically investigating such micro-macro coupling, such simulations are expensive, even under simplification, and resist analytical inquiry into system dependencies upon parameters and the nature of underlying mathematical structures. Continuum models that describe the suspension dynamics in terms of partial differential equations (PDEs) can provide a powerful and complementary approach to not only simulate the large-particle limit, but also to analyze stability of states and the structure of the collective dynamics \cite{hohenegger2010stability,ohm2022weakly,albritton2023stabilizing,coti2023orientation,Weady2024}. Such models can be posed through various means, for example by symmetry principles \cite{Simha2002} or through kinetic theories \cite{SS2008}. For the latter one might seek a probabilistic description through an $N$-particle distribution function which itself satisfies a $5N$-dimensional Fokker-Planck equation \cite{Skultety2020}. More typically studied is the so-called mean-field limit in which the single-particle distribution function is evolved \cite{SS2008}. This approximation explicitly considers the microscopic particle dynamics, giving a clear connection between the micro- and macroscopic scales, and coarse-grains hydrodynamic interactions. Justifying this procedure requires assumptions about particle separation as close interactions can be complex and nearly singular \cite{Duerinckx2024}. Nonetheless, such mean-field kinetic theories are useful and predictive for experiments, explaining transitions from isotropy to large-scale flows as system parameters, such as the swimmer volume fraction, are varied \cite{Peng2021,liu2021density}, and suggesting system-size flow structures should be dominant. 

Important issues remain. Unresolved are the nonlinear nature of the long-time dynamics, and to what extent mean-field theories, based as they are on dilute suspension assumptions, can describe the full dynamics of large-scale active swimmer suspensions. We explore these questions here. First, we simulate very large suspensions of slender self-propelled particles. Hydrodynamic interactions are described by slender body theory, which relates the velocity and force distributions along each particle through a local anisotropic mobility matrix \cite{batchelor1970slender,SS2007,SS2012}. Using a Fast Multipole Method \cite{yan2021kernel} and implicit resolution of steric interactions via a complementarity formulation~\cite{yan2019computing,Yan2020}, we analyze the limiting behavior as the adimensional particle size tends to zero at fixed volume fraction.  Performing simulations of up to nearly 20 million particles allows empirical assessment of the mean-field limit. At low volume fractions, large-scale flow quantities and correlation functions approach values independent of the particle size relative to the domain size. These results compare favorably with predictions and simulations of the mean-field PDEs, which suggest the emergence of a fluctuating flow structure whose scale is set by the simulation box size. Moving to higher volume fractions, these two descriptions diverge. Within the particle simulations, the scaling with box size disappears and the correlation length instead scales with the particle size relative to the box size, thus tending to zero and reflecting a breakdown of a dilute mean-field limit. The observed long-time particle flows are now cellular in structure, with the flow-cell size reflecting an emerging hydrodynamic screening length which scales on particle length. This is not captured by the dilute mean-field PDEs.

\begin{figure}[t!]
    \centering
    \includegraphics[width=\linewidth]{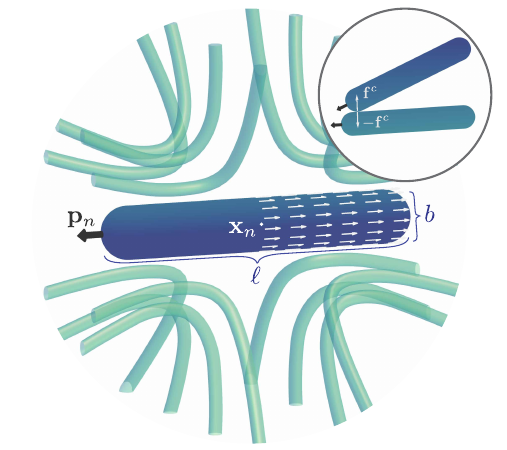}
    \caption{Schematic of the micro-mechanical model. Each particle is a spherocylinder of diameter $b$ and length $\ell$. The particle configuration is determined by the center of mass positions $\bx_n \in V$ and orientations $\bp_n \in S$, $n = 1, \ldots, N$. A tangential slip velocifty (white arrows) is prescribed on the particle surface. The consequent flows (green streamlines) and hydrodynamic forces are computed using a local slender body approximation, while steric interactions (inset in the top right) are treated using a constraint-based collision resolution algorithm.}
    \label{fig:schematic}
\end{figure}

\section{Active suspension modeling}\label{sec:discrete-model}

\subsection{The micro-mechanical model}

We consider suspensions of $N$ spherocylindrical self-propelled particles, each of length $\ell$ and diameter $b = \ell/5$, in a triply periodic domain $V = [0, L]^3$ with linear dimension $L$. The $n^{th}$  particle is represented by its center of mass position $\bx_n(t) \in V$ and orientation vector $\bp_n(t) \in S$, where $S = \{ \bp \in \mathds{R}^3 : |\bp| = 1\}$ is the unit sphere of orientations. Swimmers move with translational and rotational velocities $\dot{\bx}_n(t)$ and $\dot{\bp}_n(t)$, respectively, determined by their propulsive actuation, the hydrodynamic influence of other swimmers, and steric forces and torques $\bF_n^c$ and $\bT_n^c$ induced by particle-to-particle collisions. Their self-propulsion is induced by a prescribed slip velocity $u_n^s(s)$, with $|s|\leq\ell/2$ the signed arclength along the swimmer centerline, specified over a half of the body (similar to squirmers \cite{Lighthill1952,Blake1971,Pedley2016}). In particular, we choose $u_n^s(s) = 2U[H(s)-1]$, where $H(s)$ is the Heaviside function and $|U|$ is the isolated speed of a single swimmer. We restrict our attention to ``pusher'' particles, $U > 0$, for which the dynamics are found to be nontrivial. Following \cite{SS2007, SS2012}, the total force balance along the particle centerline is:
\begin{equation}
\dot{\bx}_n + s\dot{\bp}_n + u_n^s(s) \bp_n - \bu_n(s) = \eta(\bI + \bp_n\bp_n)\cdot\bff_n(s),\label{eq:sbt_balance}
\end{equation}
where $\bff_n(s) = \bff_n^h(s) +  \bff_n^c(s)$ is the force distribution resulting from hydrodynamic and steric forces, and $\eta = \log(2\ell/b)/4\pi\mu$ is the slender body coefficient with $\mu$ the dynamic viscosity. Here $(\bp_n\bp_n)_{ij} = p_{n,i}p_{n,j}$ is the outer product of $\bp_n$ with itself and the dot denotes tensor contraction.

The operator $\eta\left(\bI + \bp_n\bp_n\right)$ multiplying the force is the local mobility matrix that arises from shape anisotropy. The hydrodynamic force distribution and the resulting induced fluid flow $\bu_n(s) = \bu(\bx_n + s\bp_n)$ are unknown whereas the steric distribution is set by the induced body forces and torques: $\bff_n^c(s) = (1/\ell) \bF_n^c + (12 s / \ell^3) \bT_n^c \times \bp_n.$ Using the force- and torque-free conditions 
$[\bff_n(s)]_\ell = \bF_n^c$ and $[ s\bp_n\times\bff_n(s) ]_\ell = \bT_n^c$, where $[ \cdot ]_\ell = \int_{-\ell / 2}^{\ell / 2} \cdot ~ ds$ denotes an integral over the particle centerline. The rigid body dynamics are then
\begin{align}
    \dot\bx_n &= U\bp_n + \frac{1}{\ell}[\bu_n(s)]_\ell + \frac{\eta}{\ell}(\bI + \bp_n\bp_n)\cdot \bF_n^c,\label{eq:xdot_n}\\
    \dot\bp_n &= \frac{12}{\ell^3}(\bI-\bp_n\bp_n)\cdot[ s\bu_n(s)]_\ell + \frac{12\eta}{\ell^3} \bT_n^c\times\bp_n.\label{eq:pdot_n}
\end{align}
On physical considerations, we set $U = \beta\ell$, where $\beta$ is an $O({\rm s}^{-1})$ constant.

In the slender body formulation, the fluid velocity is determined by the Stokes equations forced by a singularity distribution along the centerline of each particle,
\begin{gather}
    -\mu \Delta \bu + \grad q = \sum_{n = 1}^N [ \bff_n(s) \delta_{\bx_n + s\bp_n} ]_\ell, \label{eq:stokes_a}\\
    \nabla\cdot\bu = 0,\label{eq:stokes_b}
\end{gather}
where $q$ is the pressure and we use the notation $\delta_{\bq_n} = \delta(\bq - \bq_n)$ for the Dirac delta with the domain of definition implicitly assumed. From a computational perspective, these equations can be solved efficiently in terms of the fundamental solution (i.e. the Stokeslet) coupled with the Fast Multipole Method. A schematic of the particle model and its self-generated flow is shown in Fig. \ref{fig:schematic}, where the characteristic flow field has the structure of that induced by a force dipole.

Resolving the flow between two nearly-touching particles requires prohibitively fine spatial and temporal resolution. Here, instead of resolving these close interactions, we employ a contact resolution algorithm that enforces a no-overlap condition at every time step under the assertion that particles that are not in contact exert no contact forces on one another, and those in contact exert equal and opposite forces that obey D'Alembert's principle. This approach overcomes the numerical stiffness associated with potential-based methods at the cost of an iterative solution. Details on the constraint-based formulation and its implementation can be found in the Appendix.

For a fixed particle geometry and effective volume fraction ${\nu = N\ell^3 / L^3}$, the free physical parameters are the particle size $\ell$, the scale $\beta$ of the isolated swimmer velocity $U=\beta \ell$, the viscosity $\mu$, and the domain size $L$. Non-dimensionalizing by the domain size $L$, propulsive time scale $1 / \beta$, and force per unit volume scale $\mu U / L^2$, the only remaining parameter is the dimensionless particle size $\ell' = \ell / L$ or, equivalently, the relative domain size $L' = L / \ell$. 

\subsection{Empirical distribution}

The particle configuration can be compactly described by the empirical distribution in position and orientation space,
\begin{equation}
    \Psi_N(\bx, \bp, t) = \frac{|V|}{N} \sum_{n = 1}^N \delta_{\bx_n, \bp_n},\label{eq:Psi_N}
\end{equation}
where $(\bx, \bp) \in V \times S$. Defining $[f]_S = \int_S f \Psi_N ~ dS_\bp$, this distribution yields the order parameters $c = [1]_S$, $\bn = [\bp]_S$, and $\bQ = [\bp \bp - \bI / 3 ]_S$, which are the concentration, polarity, and nematic tensor, respectively. These order parameters are formally singular for finite $N$, but we are interested in their limiting behavior as $N \rightarrow \infty$. We study this limit with the spatial autocorrelation function
\begin{equation}
    \Corr[\Phi](r) = \frac{1}{|S|} \int_S \cF^{-1}\big[|\tilde\Phi|^2 \big ](r \hat\br) ~ dS_{\hat\br},
\end{equation}
where $\tilde\Phi(\bk) = \cF[\Phi](\bk)$ denotes the spatial Fourier transform of the field $\Phi$. Correlation functions of vector and tensor valued functions are summed over each component. The volume-weighted $L^2$ norm can be determined from the correlation function by $\norm{\Phi}_2^2 = \lim_{r\rightarrow 0} \Corr[\Phi](r)$. In practice, the Fourier coefficients are computed using a Non-Uniform Fast Fourier Transform \cite{Barnett2019}; further details can be found in Appendix \ref{sec:corr}.

\section{The mean-field limit}\label{sec:mean-field}

The micro-mechanical model naturally leads to a continuum theory in the so-called mean-field limit $\ell/L\rightarrow 0$ and $N\sim (L / \ell)^3 \rightarrow\infty$. This theory, building on the Doi theory for polymer solutions \cite{Doi1986}, was first introduced and analyzed in \cite{SS2008}. We summarize the model here; a formal derivation is given in Appendix \ref{sec:mean_field}, while a rigorous derivation with precise error bounds can be found in Ref. \cite{Duerinckx2024}. 

\subsection{Constitutive equations}

For $N\rightarrow\infty$, the limiting empirical distribution $\Psi_N\rightarrow\Psi$, defined in Eq. (\ref{eq:Psi_N}), satisfies the Smoluchowski equation in position-orientation space,
\begin{equation}
    \frac{\partial\Psi}{\partial t} + \grad_\bx\cdot(\dot\bx\Psi) + \grad_\bp\cdot(\dot\bp\Psi) = 0,\label{eq:dpsi/dt}
\end{equation}
where $\grad_\bx$ denotes the spatial gradient operator and $\grad_\bp$ the gradient operator on the unit sphere. Unlike Ref. \cite{Skultety2020}, because the dynamics are deterministic, this distribution entirely characterizes the suspension and does not require an $N$-particle distribution function. That is, the characteristics of the Smoluchowski equation are precisely the discrete dynamics for all $N$ \cite{Golse2016}. Assuming particles are well-separated and taking $\ell / L \rightarrow 0$, the configuration fluxes take the simple form
\begin{equation}
    \dot\bx = \bu,\quad
    \dot\bp = (\bI - \bp\bp)\cdot(\bp\cdot\grad\bu),\label{eq:qdot}
\end{equation}
where the convention $(\grad\bu)_{ij} = \partial u_j/\partial x_i$ is used so that $(\bp\cdot\grad\bu)_i = p_j\partial u_i/\partial x_j$. Note that, in comparison to the kinetic theory of \cite{SS2008}, there is no contribution from self-propulsion when $\ell/L\rightarrow 0$ as this term is $O(\ell/L)$ due to the scaling $U \sim \ell$. 

A constitutive relation for the fluid velocity appearing in the configuration fluxes above can be derived using Batchelor's formula \cite{Batchelor1970} and is given by
\begin{gather}
    -\mu\Delta\bu + \grad q = \div( \sigma_a\bQ + \sigma_b\bS : \bE ),\label{eq:Stokes}\\
    \div\bu = 0.\label{eq:divu}
\end{gather}
Here, $\sigma_a = -\beta\nu/8\eta$ is the dipole coefficient, $\sigma_b = \nu/24\eta$ characterizes the constraint stress arising from particle rigidity, and $\bE = (\grad\bu + \grad\bu^T) / 2$ is the symmetric rate of strain tensor. The double dot notation denotes tensor contraction, $(\bS : \bE)_{ij} = S_{ijk\ell}E_{k\ell}$. We refer to equations (\ref{eq:dpsi/dt})-(\ref{eq:divu}) as the dilute mean-field limit, though we drop the qualifier where clear. 

\begin{figure*}[t!]
    \centering
    \includegraphics[width=\linewidth]{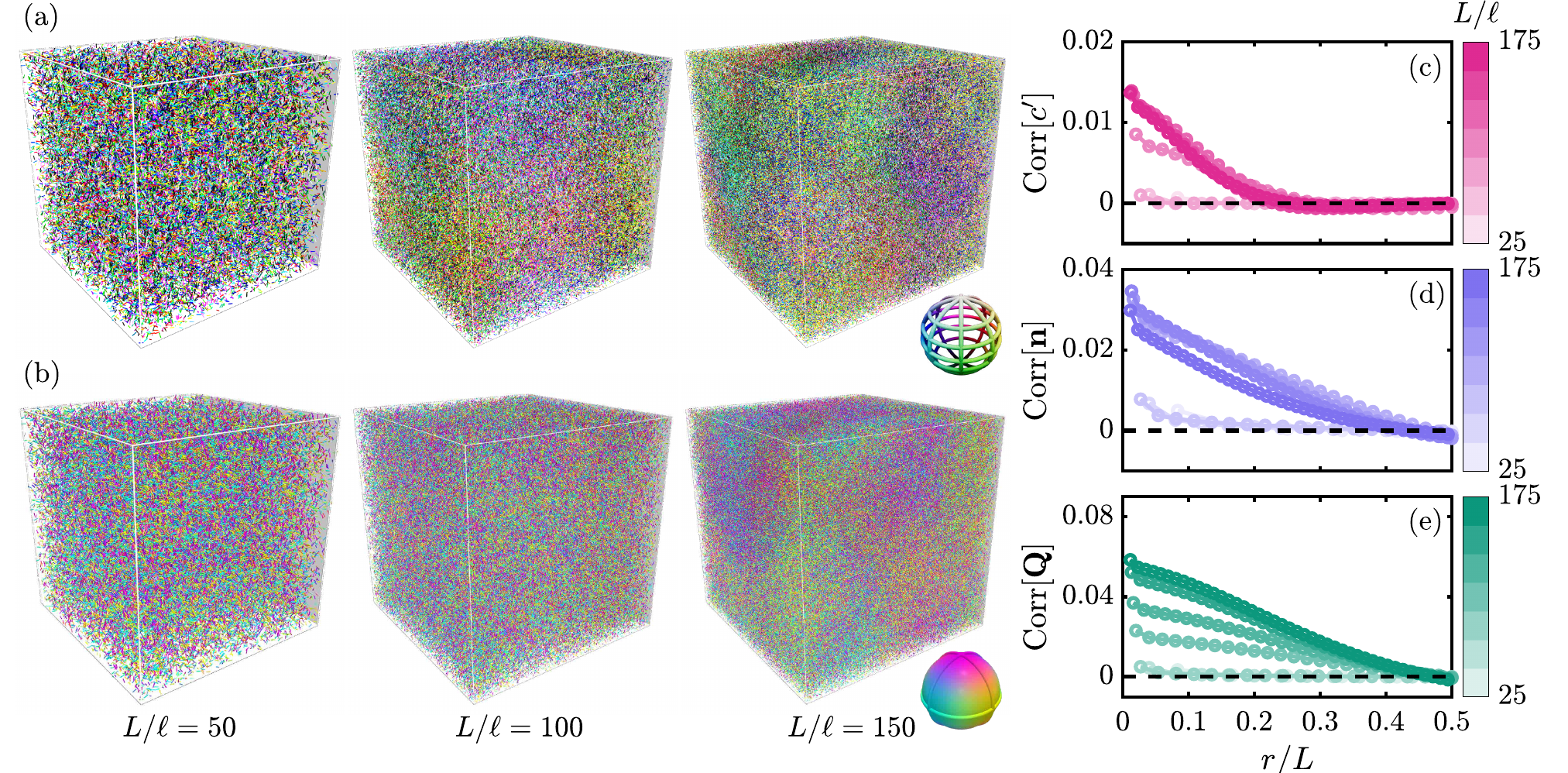}
    \caption{Orientation structures span system size in dilute suspensions ($\nu = 0.625$). Panels (a) and (b) show snapshots of particles colored by their polar orientation and nematic orientation according to the inset color maps, respectively, for $L/\ell = 50, 100$, and $150$. The right column shows the correlation functions of the (c) concentration fluctuations $c' = c - 1$ (d) polarity and (e) nematic tensor for increasing domain sizes (light to dark). At intermediate domain sizes, the magnitude of concentration and polarity correlations depends weakly on the domain size while the nematic correlation function increases in magnitude.}
    \label{fig:fig2}
\end{figure*}

We non-dimensionalize these equations by the domain size $L$ and time scale $t_c = \mu / |\sigma_a|$. (Note the time scale here differs from the particle simulations by the dimensionless factor $\nu / 8\mu\eta$.) This leaves $\zeta = \sigma_b / \mu$ as the sole dimensionless parameter which depends only on the volume fraction (linearly) and the particle aspect ratio. In the remainder of this Section we assume all variables are dimensionless.

\subsection{Analytical predictions}

One immediate observation of the mean-field limit is that the domain size is the only length scale. Indeed, the equations are invariant under the rescaling $\bx \mapsto a\bx$. This lack of an additional length scale is further made evident through linear stability analysis, in which the growth rate $\gamma$ of perturbations about the isotropic state is independent of the wavenumber \cite{ezhilan2013instabilities}. Specifically, taking $\Psi = \Psi_0 = 1/4\pi$ as the steady state solution and perturbing $\Psi = \Psi_0 + \varepsilon \tilde \Psi({\bk}, \bp) \exp(\gamma t + i\bk\cdot\bx)$, we find
\begin{equation}
    \gamma = \frac{1}{5}\left(\frac{1}{1 + \zeta/15}\right).
\end{equation}
In dimensional form, because the characteristic time scale is $t_c \propto 1 / \nu$ and $\zeta \propto \nu$, this implies the growth rate increases monotonically with $\nu$ but is bounded from above.

At the nonlinear level, particle alignment can be characterized by the relative configurational entropy \cite{SS2008,Weady2024},
\begin{equation}
    \cH[\Psi](t) = \int_V \int_S \Psi \log\left(\frac{\Psi}{\Psi_0}\right) ~ dS_{\bp} dV_{\bx}.
\end{equation}
This functional is non-negative, vanishing only on the isotropic state $\Psi = \Psi_0$. Using the Smoluchowski and Stokes equations, it is straightforward to show $\cH$ satisfies
\begin{equation}
    \frac{d\cH[\Psi]}{dt} = 3\int_V (2\bE : \bE + \zeta \bE : \bS : \bE) ~ dV_{\bx}.\label{eq:Hdot}
\end{equation}
Since $\bE : \bS : \bE \geq 0$, the relative entropy is non-decreasing and constant only when $\bE = \bm 0$, for which $\Psi$ is spatially constant. This suggests particles become more ordered over time. Note that we've assumed particles are extensile, that is $\beta > 0$. Contractility simply changes the sign of the right hand side of Eq. (\ref{eq:Hdot}). Moreover, the time scale $t_c \propto  1 / |\sigma_a|$ shows, in dimensional  terms, that the relative configurational entropy grows faster as $|\sigma_a|$ increases.

While the relative entropy can increase without bound, the fluid velocity remains bounded. Indeed, dotting the Stokes equation with $\bu$ and integrating over the fluid volume $V$ gives
\begin{equation}
    \int_V 2\bE : \bE + \zeta\bE : \bS : \bE ~ dV_{\bx} = \int_V \bE : \bQ ~ dV_{\bx}.
\end{equation}
Applying the Cauchy-Schwarz inequality, we find $\norm{\bE}_2 \leq \norm{\bQ}_2/2 \leq \sqrt{1 / 6}$, under the assumption $\trace(\bQ) = 1$, which holds for all time if the initial data satisfies $\trace(\bQ_0) = 1$. The Poincar\'e inequality $\norm{\bu}_2 \leq (1 / 2\pi)\norm{\bE}_2$ then implies the uniform-in-time bound, independent of $\zeta$,
\begin{equation}
    \norm{\bu}_2 \leq \sqrt{\frac{1}{24 \pi^2}}\label{eq:u_bound}.
\end{equation}
Note that, based on the non-dimensionalization, this implies the dimensional velocity norm grows linearly with the domain size $L$. We assess these predictions, as well as numerical simulation of the dilute mean-field model, in the following section.

\section{Dilute suspensions}\label{sec:dilute}

The dilute regime refers to suspensions with low volume fractions (typically $\nu \lesssim 1$; see Ref. \cite{Doi1986}), where particles are well separated and forces are primarily hydrodynamic rather than steric. In this section, we simulate dilute suspensions with effective volume fraction $\nu = 0.625$ (true volume fraction $\approx 2\%$)and relative domain sizes in the range $L / \ell \in [25,175]$, ranging from $O(10^4)$-$O(10^6)$ particles. Each simulation begins with a configuration of particles that is uniformly distributed in both space and orientation and is run until the velocity and configurational statistics appear to be well-sampled after the initial transient. We compare these results to simulations of the mean-field limit (\ref{eq:dpsi/dt})-(\ref{eq:divu}).

\subsection{Correlations}

Figure \ref{fig:fig2}(a)-(b) shows snapshots of the polar orientation vector $\bp$ and nematic orientation tensor $\bp\bp$ for relative domain sizes $L/\ell = 50, 100, 150$. As $L / \ell$ increases, the spatial extent of aligned regions grows, evident from the large areas of nearly uniform color, particularly in the nematic orientation shown in panel (b). The characteristic size of correlated regions is quantified using time-averaged autocorrelation functions, $\overline{\Corr[c]}, \overline{\Corr[\bn]}$, and $\overline{\Corr[\bQ]}$, where the time average is defined as $\overline{\Phi(t)} = \lim_{T\rightarrow\infty}(1 /T) \int_0^T \Phi(t') ~ dt'$. These correlation functions are plotted in panels (c)-(e) for increasing relative domain size $L / \ell$. 

The correlation length for each quantity is estimated as the first zero-crossing of the correlation function. For the polar and nematic order parameters [panels (d) and (e)], this crossover point extends to approximately half the domain, while for the concentration field [panel (c)], it reaches about one-fourth of the domain size. Correlation lengths that reach half the box size in a periodic domain indicate that the correlation length likely exceeds the largest physical length scale permitted by the triply periodic boundary conditions. Notably, the amplitude of the nematic correlation function increases with the domain size, consistent with the mean-field model’s prediction of increasing relative entropy $\cH[\Psi]$. 

\begin{figure*}[t!]
    \centering
    \includegraphics[width=\linewidth]{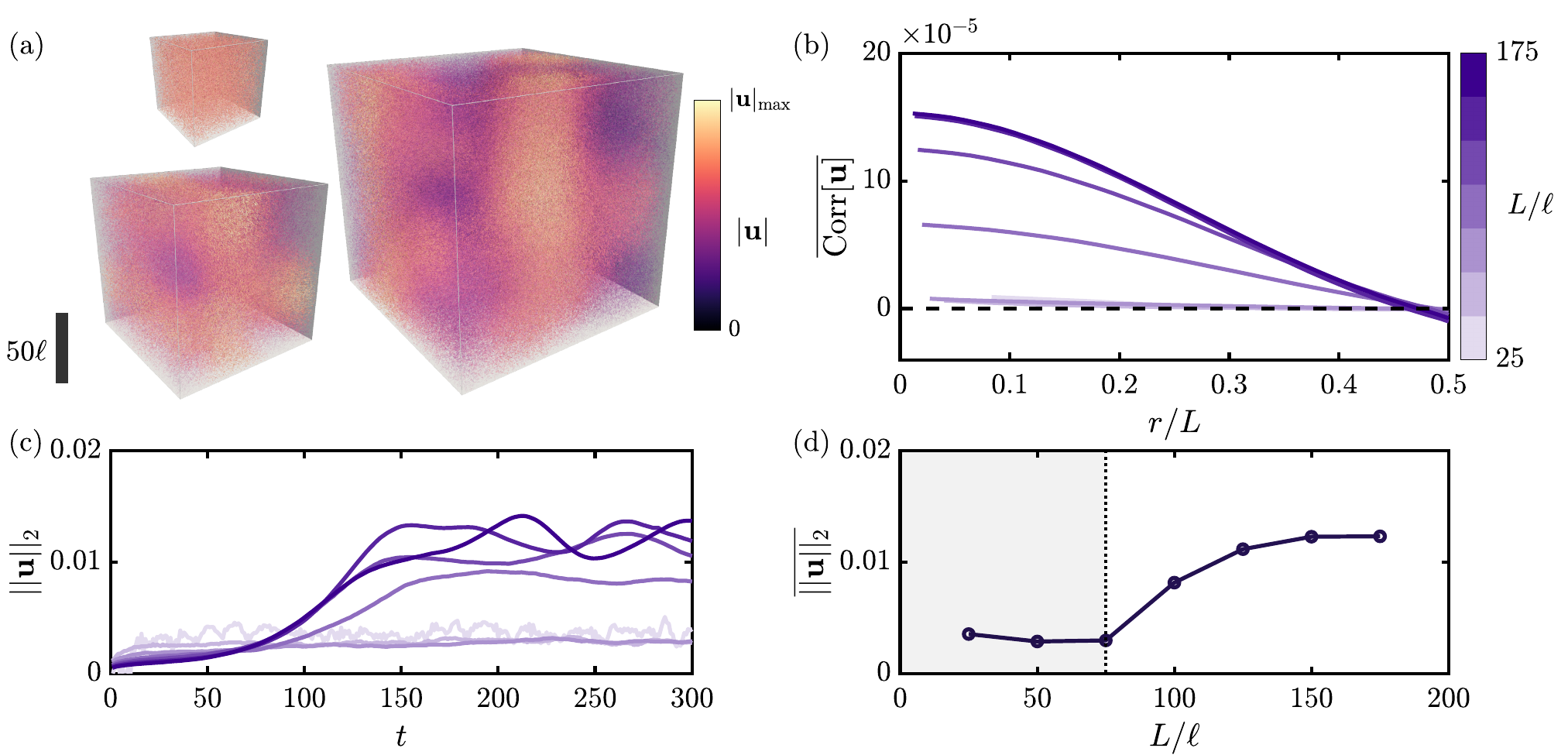}
    \caption{Convergence of dilute suspensions ($\nu = 0.625$) in the mean-field limit. (a) The magnitude of the velocity shows collective flows span the entire domain. The scale bar denotes approximately 50 particle lengths. (b) The time-averaged velocity autocorrelation function converges to a universal curve whose correlation length is half the domain size. (c) The growth rate and overall magnitude of the velocity norm approach similar values as $L / \ell\rightarrow\infty$. (d) Time average of the normalized velocity norm at the quasi-steady state ($t = 200$--$300$). Following the transition to collective motion (dotted line), the averaged norm gradually increases with $L / \ell$, appearing to converge to a finite value with $\overline{\norm{\bu}_2} \approx 0.012$. A movie of the simulation with $L = 150$ is available as Supplementary Material.}
    \label{fig:fig3}
\end{figure*}

System-scale correlations are even more pronounced in the velocity field, as illustrated in Figure \ref{fig:fig3}. Panel (a) shows snapshots from the same simulations depicted in Figure \ref{fig:fig2}, where particles are now colored by their speed. The velocity field forms extensive regions of high- and low-speed flow. These large scale flows are examined by the time-averaged velocity autocorrelation function $\overline{\Corr[\bu]}$, shown in panel (b). As $L / \ell$ increases, the velocity autocorrelation function appears to approach a single curve, with nearly indistinguishable results for $L / \ell = 150$ and $175$. 

Figures 3(c) and (d) further illustrate the convergence of velocity and time scales. Panel (c) shows the evolution of the velocity norm $||\bu||_2$, showing instability of the isotropic state with a growth rate that appears to converge as $L/\ell$ increases. Following this initial instability, the velocity remains bounded, consistent with the analytical bound (\ref{eq:u_bound}). Fluctuations about the long-time mean exhibit similar time scales and magnitudes. Panel (d) shows the time average of the velocity norm over the fluctuating region, indicating that the mean long-time value is also convergent for large $L/\ell$. The dotted line in Panel (d) represents an empirical threshold for instability. This threshold, whose dependence on the domain size at finite values of $\ell / L$ is predicted by kinetic theory \cite{SS2008}, arises as a consequence of particle motility.

It is interesting to note that, while the velocity correlation function appears to converge, the nematic orientation correlation function continues to increase over the same domain sizes. This observation is consistent with the analytical results of \cite{Hofer2024, Duerinckx2024} in which the mean-field velocity is described by the mean-field model to $O(\nu)$ while the distribution is only described to $o(\nu)$.

\begin{figure*}[t!]
    \centering
    \includegraphics[width=\linewidth]{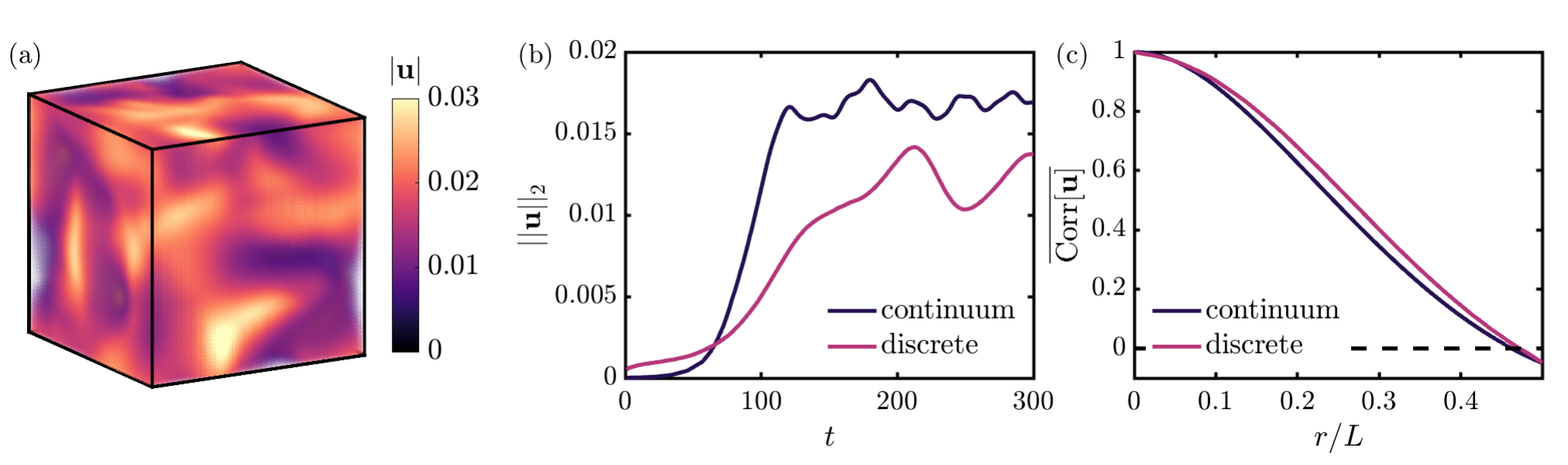}
    \caption{Numerical simulation of the mean-field limit. (a) A snapshot of the velocity magnitude at a late time in the simulation shows structures on the scale of the domain, similar to the discrete model. (b) Temporal evolution of the velocity norm compared to the discrete model  ($L / \ell = 175$). The mean-field limit transitions to collective motion faster and has a larger long-time mean, though the overall scales are comparable. (c) The correlation function of the velocity field closely follows that of the discrete model.}
    \label{fig:fig4}
\end{figure*}

\subsection{Comparison to the mean-field limit}

Convergence of the velocity correlation function is consistent with convergence to a mean-field limit. We can assess this more directly by simulating the mean-field model. For this, we use quasi-equilibrium closure approximation, often referred to as the Bingham closure, which evolves only the second orientational moment $\bQ$ of the particle distribution function, thereby eliminating orientational degrees of freedom \cite{chaubal1998closure,gao2017analytical,Weady2022,Weady2022b}. Higher order orientational moments are then approximated in terms of the Bingham distribution $\Psi_B(\bx, \bp, t) = Z^{-1} \exp(\bB:\bp\bp)$, where the normalization factor $Z(\bx, t)$ and exponent $\bB(\bx, t)$ are determined from the constraint $\int_S \bp\bp\Psi_B ~ dS_{\bp} = \bQ$ at each point in space. The dynamical equation for $\bQ$ is explicitly
\begin{equation}
    \bQ_t + \bu\cdot\grad\bQ - (\grad\bu\cdot\bQ + \bQ\cdot\grad\bu^T) + 2\bS_B : \bE = \bm 0,    
\end{equation}
where $\bS_B = \int_S \bp\bp\bp\bp \Psi_B ~ dS_{\bp}$ is the fourth moment of the Bingham distribution. The constraint stress in the Stokes equations is similarly approximated by $\zeta\bS_B : \bE$. Like the kinetic theory, the Bingham closure has a configurational entropy $\cH[\Psi_B]$ whose evolution is governed by Eq. (\ref{eq:Hdot}). The numerical implementation is based on the fast solution methodology described in Ref. \cite{Weady2022}; additional details can be found in the Appendix.

\begin{figure*}[t!]
    \centering   \includegraphics[width=\linewidth]{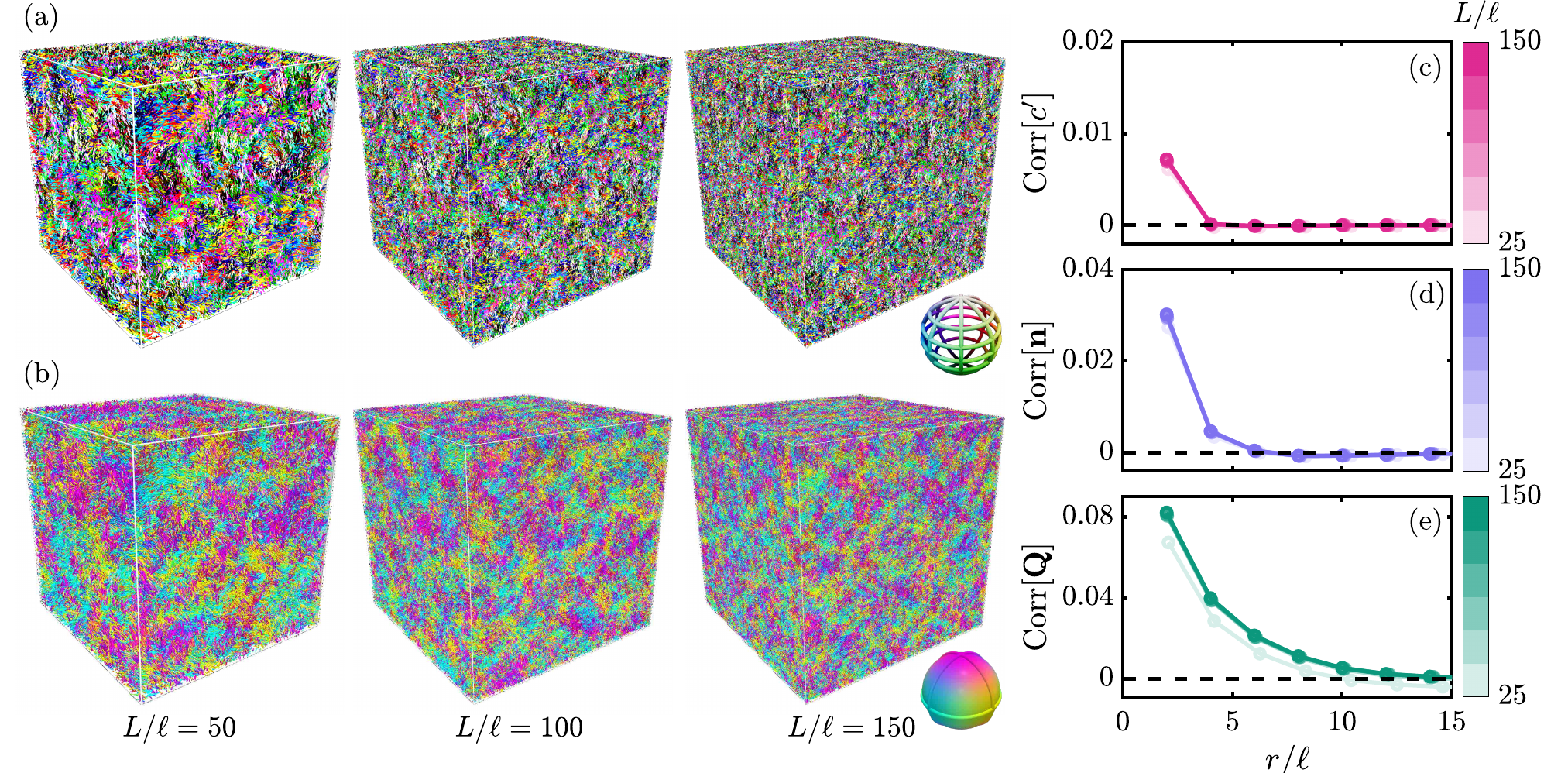}
    \caption{Orientation structures tend to zero in concentrated suspensions ($\nu = 5$). Panels (a) and (b) show snapshots of particles colored by their polar orientation and nematic orientation according to the inset color maps, respectively, for $L / \ell = 50, 100$, and $150$. The right column shows the correlation functions of the (c) concentration (d) polarity and (e) nematic tensor for increasing relative domain sizes $L / \ell$ (light to dark) as a function of the distance relative to the particle size $r / \ell$. Scaled in this way, the correlation functions appear to be independent of $\ell / L$ and thus scale to zero as $\ell / L\rightarrow 0$.}
    \label{fig:fig5}
\end{figure*}

Figure \ref{fig:fig4} shows a snapshot of the velocity field from a simulation with the dimensionless constraint stress coefficient $\zeta = 0.1421$, which corresponds, without fitting, to the particle simulations and is the only free parameter in the model. Time is rescaled by the factor $\nu / 8\mu\eta$ for consistency. The velocity field has a similar large-scale structure, with broad regions of high and low flow speed. The evolution of the velocity norm, shown in panel (b), shows a similar time scale of instability and a long-time mean to the discrete simulations, however fluctuations about this mean occur on a faster time scale. While time scales may differ, the time-averaged velocity autocorrelation in panel (c) closely follows that of the particle simulations. 

\section{Concentrated suspensions}

At higher volume fractions, $\nu \gtrsim 1$, steric forces become important and can be comparable in magnitude to hydrodynamic forces \cite{Doi1986}. This regime challenges rigorous analysis of the mean-field limit, though phenomenological models of steric interactions have been able to capture many aspects of concentrated dynamics such as the isotropic-nematic phase transition \cite{ezhilan2013instabilities}. In this section we perform simulations at a volume fraction well into the concentrated regime, $\nu = 5$ (true volume fraction $\approx 16\%$), ranging over relative domain sizes $L / \ell \in [25,150]$. For the largest domain size $L = 150$, the simulation contains approximately 16 million particles.

\begin{figure*}[t!]
    \centering
    \includegraphics[width=\linewidth]{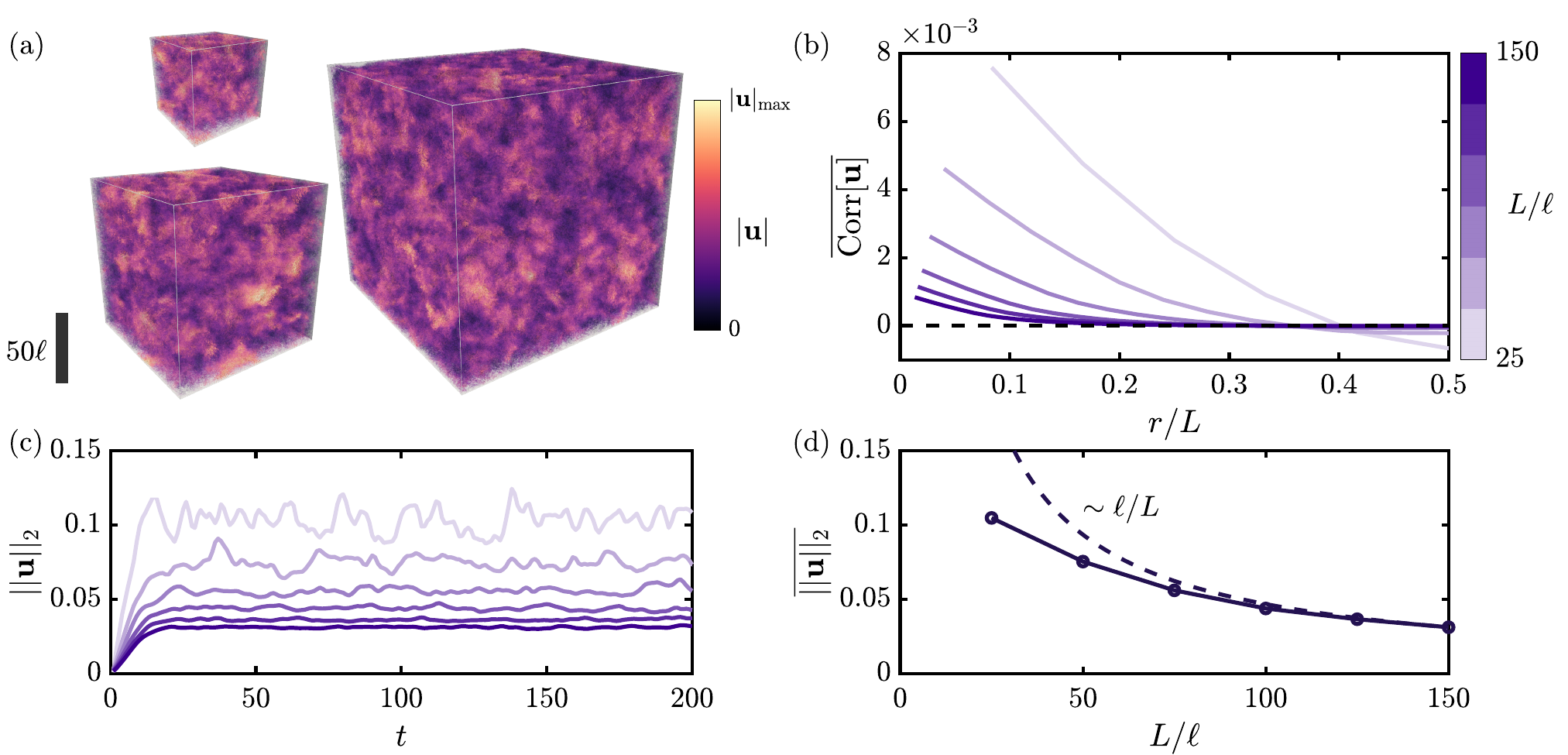}
    \caption{Non-convergence of concentrated suspensions ($\nu = 5$) in the mean-field limit. (a) The magnitude of the velocity shows fine-scale structures that scale with the particle size. The scale bar denotes approximately 50 particle lengths. (b) The time-averaged velocity autocorrelation function decreases with $L / \ell$, tending to zero in the mean-field limit. (c) The growth rate and overall magnitude of the velocity norm decrease as $L / \ell\rightarrow\infty$, as do fluctuations about the long-time mean. (d) The time average of the normalized velocity norm at the quasi-steady state ($t = 100$--$200$) decreases with $L / \ell$, showing an asymptotic scaling of $\ell / L$. A movie of the simulation with $L = 150$ is available as Supplementary Material.}
    \label{fig:fig6}
\end{figure*}

\subsection{Correlations}

Figure \ref{fig:fig5} shows snapshots of the polar and nematic orientation from simulations with $L / \ell = 50, 100$, and $150$, analogous to that displayed in Fig. \ref{fig:fig2}. Fine-scale structures emerge, and these structures remain on the scale of the particle as the relative domain size increases. This is quantified again by the autocorrelation functions, which are here displayed as functions of $r / \ell$ relative to the particle length. Unlike the dilute case, the correlations are short range, extending over less than 15 particle lengths independently of the relative domain size. The amplitude of the correlation function does not vary significantly, indicating that particles are aligned to a similar degree at the local level in all cases.

This behavior is even more pronounced in the velocity field, snapshots of which are shown in Figure \ref{fig:fig6}(a). The time-averaged correlation function $\overline{\Corr[\bu]}$ in panel (b) has a characteristic length and amplitude that rapidly decreases with increasing $L / \ell$. The velocity norm, whose time evolution is shown in panel (c), demonstrates similar trends, where the long-time mean, its fluctuations, and growth rate of instability all decrease with increasing $L / \ell$. The time averaged velocity norm shows an asymptotic $\ell / L$ scaling, indicating a trend to a no-flow state as $L / \ell \rightarrow \infty$. Such flows are not described by the dilute mean-field model, where isotropic suspensions are linearly unstable for any value of the dimensionless volume fraction parameter $\zeta$.

\begin{figure}[t!]
    \centering
    \includegraphics[width=\linewidth]{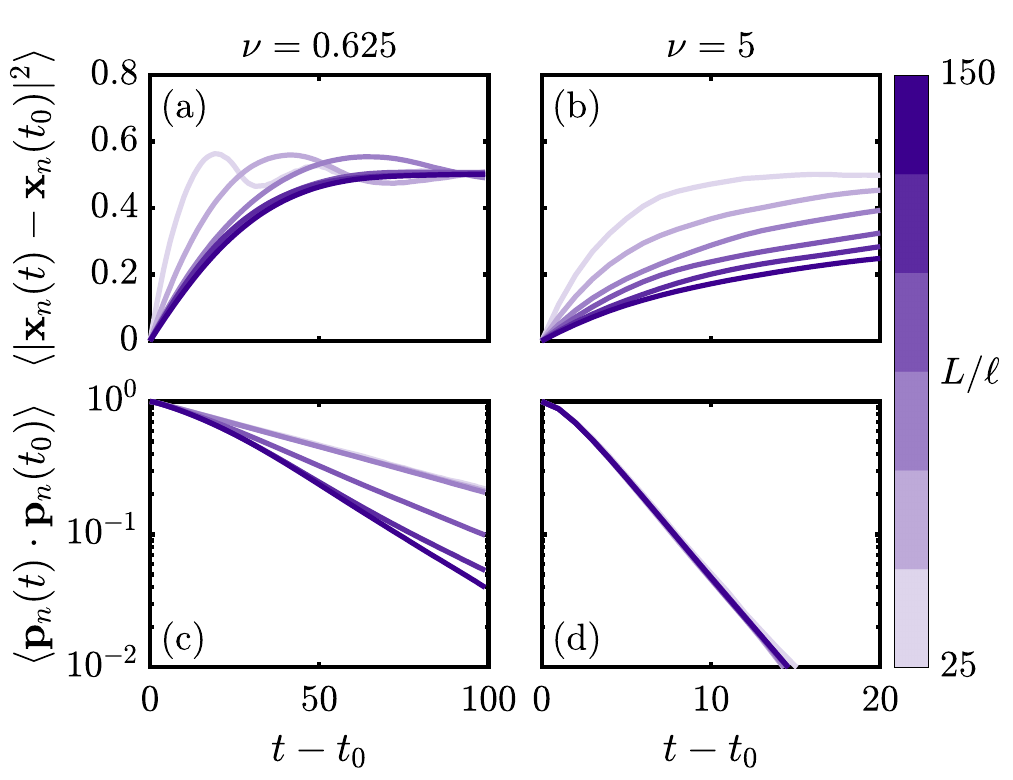}
    \caption{Mean squared displacement of the center of mass (a)-(b) and temporal orientation correlation (c)-(d) for a dilute ($\nu = 0.625$, left) and concentrated ($\nu = 5$, right) suspension. In the dilute case, both the spatial and orientational quantities appear to converge to a non-zero value, while only the orientational correlation converges in the concentrated case.}
    \label{fig:fig7}
\end{figure}

The vanishing velocity field does not necessarily imply particles are stationary. We quantify particle motion through the mean squared displacement of the center of mass $\overline{\langle |\bx_n(t) - \bx_n(t_0)|^2 \rangle}$ and the temporal orientation correlation $\overline{\langle \bp_n(t)\cdot\bp_n(t_0)\rangle}$, where $t_0$ is fixed at a time following the initial transient \cite{Doi1986,SS2007} and $\langle\cdot\rangle$ denotes an average over all particles. These quantities are displayed in Fig. \ref{fig:fig7} for both dilute ($\nu = 0.625$) and concentrated ($\nu = 5$) suspensions. The dilute case appears to converge in both the center of mass and orientation displacements, again consistent with convergence in the mean-field limit. The concentrated case decreases consistently with $L / \ell$ in its center of mass displacements and, remarkably, to a finite non-zero value in its orientation correlations. The implications of this are striking: as $\ell / L\rightarrow 0$, particles do not move spatially on average but rotate persistently.

\subsection{Implications for the mean-field limit}

The limiting behaviors of the high volume fraction simulations have significant consequences for the application of mean-field theories. Notably, because the correlation length, velocity magnitude, and mean squared displacement decrease in proportion to $\ell / L$, the mean-field limit tends towards a state of no large-scale flow and the dynamics are purely local in space. Phenomenologically, these dynamics are characterized by independent parcels, fixed in space, within which particles rotate rapidly.

An estimate of the volume fractions for which the dilute mean-field equations hold can be understood through their formal derivation. In particular, a key step in the derivation is the Taylor expansion of the fluid velocity about the particle centerline,
\begin{equation}
    \bu_n(s) = \left.\sum_{k = 0}^\infty \frac{s^k (\bp_n\cdot\grad)^k }{k!}\bu(\bx)\right\rvert_{\bx_n},\label{eq:multipole}
\end{equation}
which is truncated at linear order. This series, however, will only converge at low volume fractions. To see this heuristically, recall the velocity field is a superposition of Stokeslets. Because the Stokeslet has a $1 / r$ singularity, in the vicinity of the particle the $k$th derivative of the velocity scales at least as $\sim k!/ r^{k+1}$, where $r$ is the distance between the target and source. Namely, this means the $k$th term in the expansion (\ref{eq:multipole}) scales as $\sim (s / r)^k$, with a volume fraction dependent pre-factor. Since $s \in [-\ell / 2, \ell / 2]$, this requires $r > \ell / 2$ to converge.

\begin{figure}[t!]
    \centering
    \includegraphics[width=\linewidth]{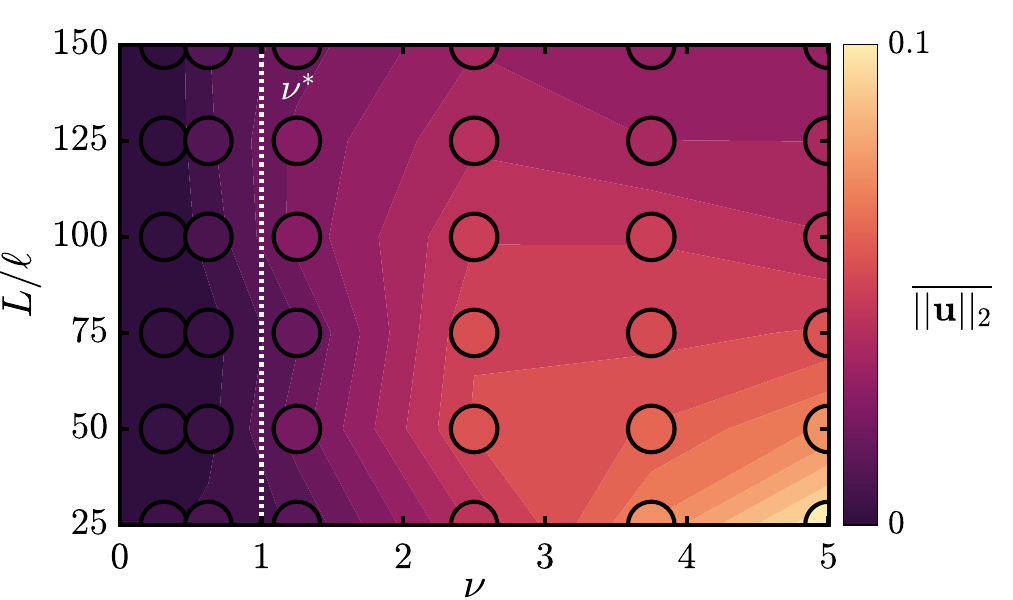}
    \caption{Phase diagram of the time-averaged velocity norm. Circles denote sampled values while the white dotted line indicates the estimated critical effective volume fraction $\nu^* = 1$ for which the dilute theory is valid.}
    \label{fig:fig8}
\end{figure}

Using this criterion, we can estimate the critical effective volume fraction $\nu^*$ for which the dilute expansion is valid by considering a uniform distribution of spheres of diameter $\ell$, reflecting a suspension of uniformly distributed rods with random orientations. For this idealized configuration, the minimum separation distance from one sphere center to the surface of its nearest neighbor is
\begin{equation}
    r_{\rm sep} = \ell\left(\frac{1}{\nu^{1/3}} - \frac{1}{2}\right),
\end{equation}
from which we conclude convergence requires $\nu < \nu^* = 1$. Physically, this formal calculation reflects a screening length: at such high volume fractions, inevitable near-body interactions will ultimately hinder flow. This threshold characterizes the transition from the dilute to concentrated regime. Figure \ref{fig:fig8} shows a phase diagram of the time-averaged velocity norm $\overline{\norm{\bu}_2}$ in the $(\nu, L / \ell)$-plane. Circles denote simulated values and the white dotted line indicates the estimated critical effective volume fraction $\nu^* = 1$. For $\nu < \nu^*$, the contours appear straight in $L / \ell$, consistent with mean-field convergence. For $\nu > \nu^*$, the contours curve as $L / \ell$ becomes large; at constant $\nu$, the norm appears to decrease monotonically.

\section{Discussion}

This work provides a comprehensive characterization of the mean-field limit of motile particle suspensions whose dynamics are strictly governed by slender body hydrodynamics and steric interactions. At low volume fractions, we found characteristic flow structures scaled with the domain size and velocity statistics converged in a manner consistent with analytical results. The spatial autocorrelation functions of both the fluid velocity and orientational order parameters reveal a correlation length at half the domain size, indicating the dominant length scale is set by the confining geometry. System-scale flow structures are seen in experiments on circularly confined bacterial suspensions, where the dominant mode is a system-spanning vortex up to millimetric scales (E. Clement, A. Lindner, B. P\'erez-Estay, {\em private communication}, May 2025). The detailed flows will differ due to confining boundary conditions, however we expect this scaling to be general.

At higher volume fractions, we found the dynamics tend towards a state with no apparent large-scale flow. This property is quantified by the spatial autocorrelation function, which, in contrast to the low volume fraction case, scales with the particle size. This scaling suggests, in the large particle number limit, the dynamics becomes localized in space and purely orientational. Moreover, the exponential decay of the orientational temporal autocorrelation function suggests orientational diffusion, driven by steric effects, dominates. It is important to note that many living systems do exhibit large-scale dynamics at high volume fractions \cite{sanchez2012spontaneous, duclos2020topological}. Our results suggest these material flows require additional effects such as cross-linking, which pose significant analytical challenges \cite{furthauer2022cross}.

It remains open whether the absence of large-scale organization persists as the volume fraction increases further. Indeed, at very high volume fractions one expects an isotropic-nematic phase transition \cite{Doi1986}, which is observed in purely steric simulations of this model with thermal fluctuations \cite{yan2019computing}. Here, activity might play the role of thermal fluctuations. The slender body representation of the particle dynamics will likely incur large errors at such volume fractions, and improvements in the micro-mechanical model, such as direct numerical simulation of the many-particle boundary value problem, will be essential.

\begin{acknowledgments}
    We thank Anke Lindner and Xiang Cheng for engaging discussions, and Wen Yan for his contributions to the implementation and derivation of the slender body particle model. B.B. acknowledges support from NSF grant AST-2009679 and NASA grant No. 80NSSC20K0500. This research was also supported in part by the National Science Foundation under Grant No. NSF PHY-1748958.
B.B. is grateful for generous support from the David and Lucile Packard Foundation, the Alfred P. Sloan Foundation, and the Flatiron Institute, which is funded by the Simons Foundation. The computations in this work were performed at facilities supported by the Scientific Computing Core at the Flatiron Institute, a division of the Simons Foundation.
\end{acknowledgments}

\vspace{0.125in}

B.P. and S.W. contributed equally to this work.

\appendix

\section{Particle simulations}

\subsection{Orientations}

Owing to the deficiencies of representing orientations using Euler angles, we avoid directly evolving $\bp$ based on $\dot{\bp}$. Rather, we follow \cite{Delong2015} and introduce the unit quaternion $\bq_n(t)=[s_n,\bw_n]^T(t)$ with scalar component $s_n$ and vector component $\bw_n$ for body $n$, chosen such that $\bp_n = \bq_n \hat{\bz}$. Here, $\hat{\bz}$ denotes the $z$-axis and $\bq_n \hat{\bz}$ is the quaternion-vector multiplication of $\bq_n$ with $\hat{\bz}$. In this form, if a body with orientation $\bq_n^k$ undergoes a rotation with constant angular velocity $\bOmega_n^k$ from time $t^k$ to time $t^k+\Delta t$, the updated orientation is given by
$
\bq_n^{k+1} = \btheta_{\Delta t \bOmega_n^k} * \bq_n^k,
$
where $\btheta_{\ba} = [\cos(\|\ba\|/2), \sin(\|\ba\|/2) \ba / \|\ba\|]^T$ is the quaternion representing a counterclockwise rotation by $\|\ba\|$ radians about an axis $\ba / \|\ba\|$ and $*$ indicates the quaternion multiplication operation.

\subsection{Contact-constrained time integration}

\begin{figure}[b!]
    \centering
    \includegraphics[width=\linewidth]{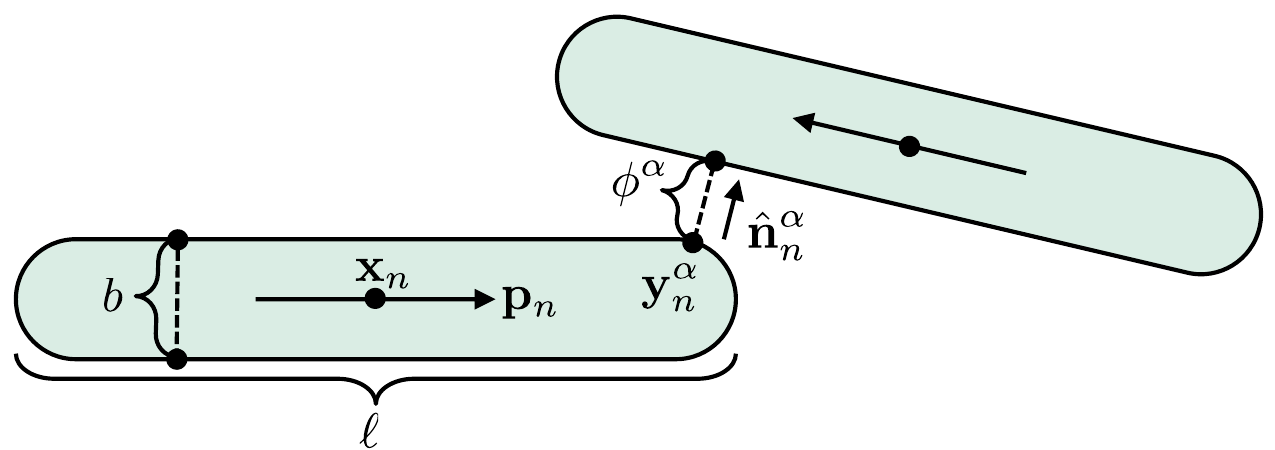}
    \caption{Contact constraint $\alpha$ between swimmers $n$ and $m$. The constraint is placed between surface points \(\by_n^{\alpha}\) and \(\by_m^{\alpha}\) on bodies \(n\) and \(m\), respectively. These points are selected such that their surface normals satisfy \(\hat{\bn}_n^\alpha = -\hat{\bn}_m^\alpha\) and that the pair-wise signed separation distance $\phi^{\alpha}$ is minimized.}
    \label{fig:contact_schem}
\end{figure}

In the presence of impulses generated by collision events, the center of mass forces $\bF^c$ and torques $\bT^c$, and the induced center of mass motions $\dot{\bx}$ and $\dot{\bq}$, may experience sudden jumps/discontinuities in time, making their time evolution unrepresentable as a classical ordinary differential equation with smooth-right hand side. Instead, the time evolution of rigid colliding bodies falls into the class of differential variational inequality (DVI) problems \cite{Pang2008}. Smooth collision resolution methods, e.g., those that rely on pairwise repulsive potentials, smooth out the effect of collisions such that the DVI simplifies down to a standard smooth ODE. Unfortunately, this smoothing process introduces inaccuracies and severe numerical stiffness that often necessitates the use of inhibitively fine timestep sizes. Nonsmooth-collision resolution methods, on the other hand, directly discretize the underlying differential variational inequality problem and have better numerical stability \cite{negrut_demp_vs_demc_2017}.

Mathematical and numerical details for this discretization process can be found in \cite{Anitescu2006} and \cite{Weady2024b}, respectively, and a summary is provided in what follows. The first component of this procedure is the discretization of the temporal dynamics using left-sided Riemann sums. Under simple assumptions about the continuity and boundedness of $\bx(t),\bq(t),\dot{\bx}(t),$ and $\dot{\bq}(t)$ \cite{Anitescu2006}, the system can then be evolved forward in time from time $t^k$ to $t^{k+1} = t^k + \Delta t$ using
\begin{equation}
\bx^{k+1}_n = \bx^{k}_n + \Delta t \dot{\bx}^k_n, 
\quad\bq^{k+1}_n = \btheta_{\Delta t \bOmega_n^k} * \bq_n^k.
\end{equation}
While this may appear identical to a first order Taylor expansion, Taylor's expansion theorem is inapplicable due to the non-smoothness of collisions. 

The second component is the representation of collision-resolution as a constrained optimization problem and the introduction of no-overlap constraints. Fig.~\ref{fig:contact_schem} illustrates the $\alpha$\textsuperscript{th} no-overlap constraint. This constraint, which acts between a pair of rods $n$ and $m$ enforces that the shared-normal signed separation distance $\Phi^\alpha$ between the pair remains positive, while also ensuring that no contact force is applied if the rods remain separated. Intuitively, $\Phi^\alpha$ represents the Euclidean distance between two points $\by^\alpha_n$ and $\by^\alpha_m$ on the surface of the opposing rods, adjusted to be negative whenever the rods overlap. The condition that these points share a normal, i.e., $\hat{\bn}^\alpha_n = -\hat{\bn}^\alpha_m$, ensures that the collision forces on opposing rods, which act along the surface normals, are equal and opposite in direction. 

Letting $\lambda^\alpha$ represent the Lagrange multiplier for the constraint, the no-overlap condition can be written explicitly as
\begin{equation}
    \begin{aligned}
    \text{No contact: }& \Phi^\alpha\geq 0, \quad \lambda^\alpha=0, \\
    \text{Contact: }& \Phi^\alpha=0, \quad \lambda^\alpha\geq0,
    \end{aligned}
   \label{eq:flip_flop}
\end{equation}
and the subsequent center of mass force and torque induced by this constraint as
\begin{equation}\label{eq:collision_ft}
   \bF_n^c = -\sum_{\alpha\in \mathcal{A}_n} \hat{\bn}^\alpha_n \lambda^\alpha, \quad 
   \bT_n^c = -\sum_{\alpha\in \mathcal{A}_n} (\by^\alpha_n - \bx_n) \times (\hat{\bn}^\alpha_n \lambda^\alpha),
\end{equation}
where $\mathcal{A}_n$ is the set of all constraints affecting particle $n$. Mathematically, Eq. \eqref{eq:flip_flop} yields a nonlinear complementarity problem, written over the set of pairs as:
$	
\boldsymbol{0} \leq \bPhi \perp \blambda \geq \boldsymbol{0},
$
where $\bPhi=(\Phi_0,\Phi_1,...)$ and $\blambda=(\lambda_0,\lambda_1,...)$ denote the collections of all minimal distances and contact force magnitudes. Here, the inequalities are applied element-wise and the perpendicular symbol $\perp$ emphasizes that the separation distance and contact force magnitude are mutually orthogonal, i.e., $\bPhi \cdot \blambda = 0$. While nonlinear complementarity problems can be solved via Newton's method, the computational cost can be steep and solutions are not guaranteed to exist \cite{erleben_siggraph_2013}. The common alternative employed in this work is to linearize $\bPhi(\blambda)$. Linearization equates to assuming that 
\begin{equation}
\begin{aligned}
    \Phi^{\alpha,k+1} &= \Phi^{\alpha,k} + \Delta t\dot{\Phi}^{\alpha,k}, \\
    \dot{\Phi}^{\alpha,k} &= -\dot{\by}_n^{\alpha,k} \cdot \hat{\bn}_n^{\alpha,k} - \dot{\by}_m^{\alpha,k} \cdot \hat{\bn}_m^{\alpha,k},
\end{aligned}
\end{equation}
or rather that the relative translational motion of the contact points along the current contact line is the only contributor to change in separation. Nonlinear contributions, such as translation along the contact plane, rotation about the contact points, or change in contact locations, are neglected. As discussed in \cite{Weady2024b}, linearization errors can be mitigated by employing surface discretizations or via adaptive constraint generation without resorting to a full Newton solve. Nevertheless, as demonstrated in \cite{yan_collision_stress_2019}, pairwise no-overlap constraints can be successfully applied to dilute rod suspensions without the need to resort to more expensive alternatives. The benefit of linearization is that the nonlinear complementarity problem can be simplified to a constrained optimization problem
\begin{equation}
    \blambda^* = \argmin_{\blambda \geq 0} \blambda^T\bPhi^k + \frac{\Delta t}{2} \blambda^T\dot{\bPhi}^k(\blambda),
\end{equation} 
the solution $\blambda^*$ of which is guaranteed to exist and is guaranteed to produce unique center of mass forces and torques \cite{pospisil_thesis_2015}.

\subsection{Slender body hydrodynamics}
The calculation of the bulk fluid velocity $\bu_n(s)$ appearing within Eq.~\eqref{eq:sbt_balance} involves solving the Stokes equations Eq.~\eqref{eq:stokes_a}-\eqref{eq:stokes_b} under the influence of the, yet-unknown, centerline force distributions $\bff_n(s)$. Slender body theory approximates this velocity by treating the hydrodynamic flow induced by each filament as a superposition of fundamental solutions to the Stokes equations. The two leading terms are the contributions from the so-called Stokeslet and doublet, which, for a source centered at $\bx_0$ and a target centered at $\bx$, are given by:
\begin{equation}
\bG(\br) = \frac{\bI + \hat\br\hat\br}{r}, \quad
\bD(\br) = \frac{\bI - 3\hat\br\hat\br}{r^3},
\end{equation}
where $r = |\br|$ and $\hat\br = \br/ r$. For a fluid element $\bx$ (i.e., a point that lies outside all rods), the fluid velocity is approximated as
\begin{equation}
    \begin{aligned}
    \bu(\bx) = 
    \frac{1}{8 \pi \mu} \sum_{n=1}^N &\biggl[ 
    \bG(\bx - (\bx_n + s\bp_n)) \cdot\bff_n(s)\biggr. \\
    +
    \frac{a_{\rm sbt}^2}{2}&\biggl.\bD(\bx - (\bx_n + s\bp_n))\cdot\bff_n(s)
    \biggr]_\ell.
    \end{aligned}
\label{eq:sbt_hydro_vel}
\end{equation}

Following \cite{Maxian2021}, we observe that $\bG(\br) + a_{\rm sbt}^2\bD(\br)/2$ is equivalent to the Rotne-Prager-Yamakawa (RPY) kernel for the hydrodynamic flow induced by a no-slip sphere of radius $a_{\rm rpy} = (\sqrt{3}/2)a_{\rm sbt}$, allowing us to efficiently evaluate Eq.~\eqref{eq:sbt_hydro_vel} using the triply periodic extension to the kernel aggregate fast multipole method implemented in the STKFMM library \cite{stkfmm2021}. Similarly, for determining the backbone surface velocity $\bu_n(s)$, rather than following past works that treat the centerline fluid velocity along a fiber as the background fluid velocity at that point (i.e. $\bu_n(s) = \bu(\bx_n+s\bp_n)$), we use the RPY sphere centerline velocity:
\begin{equation}
\bu_n(s) = \left.\left(\bu + \frac{b^2}{24} \Delta\bu\right)\right |_{\bx_n+s\bp_n},
5\end{equation}
consistent with our treatment of the fibers as a line distribution of overlapping spheres. The role of slender body theory within our simulations can then be seen as providing the anisotropic mobility matrix $\eta (\bI + \bp_n \bp_n)$ for handling self-interactions.  Note that due to our constraint-based formulation, $r \geq b$, so we need not apply regularizing corrections to the RPY kernel for overlapping particles.

By isolating the parallel and perpendicular components of the unknown centerline force density $\bff_n$ from Eq.~\eqref{eq:sbt_balance}, we obtain
\begin{align}
\eta (\bI - \bp_n\bp_n)\cdot\bff_n(s)
    &= \left(\bI-\bp_n\bp_n\right)\cdot \left(\dot{\bx}_n-\bu_n(s)\right)+s \dot{\bp}_n \\
2\eta \bp_n\cdot\bff_n(s)
    &= \bp_n\cdot\left(\dot{\bx}_n-\bu_n(s)\right) + u_n^s(s).
\end{align}
Substituting in $\dot{\bx}_n$ and $\dot{\bp}_n$ from Eq.~\eqref{eq:xdot_n} and Eq.~\eqref{eq:pdot_n}, and rearranging such that terms independent of $\bff_n(s)$ are moved to the right-hand side, produces a linear system of equations representing force-balance:
\begin{align}
&\bff_n(s) +\frac{1}{\eta}\left(\bI - \bp_n\bp_n/2\right)\cdot\bu_n(s)  \nonumber\\
&\qquad-\frac{1}{\eta\ell}\left(\bI-\bp_n\bp_n/2\right)\cdot[ \bu_n(s) ]_\ell  \nonumber\\
&\qquad-\frac{12s}{\eta\ell^3}\left(\bI-\bp_n\bp_n\right)\cdot[ s\bu_n(s) ]_\ell \nonumber\\
&\quad= \frac{1}{2\eta}\left(u_n^s(s)-U\right)\bp_n + \frac{1}{\ell}\bF_n + \frac{12s}{\ell^3}\bT_n\times\bp_n.
\end{align}
Note $\bu_n(s)$ occurs on the left-hand side because it is implicitly dependent on $\bff_n(s)$ via the slender body hydrodynamic coupling. This system can be solved by discretizing the centerline using a quadrature scheme. Once a discretization is selected, the resulting linear system must be solved using an iterative, matrix-free approach, as the number of degrees of freedom\textemdash reaching up to 810 million in our largest systems\textemdash precludes the use of an explicit matrix representation. For this purpose, we employ the Generalized Minimal Residual Method (GMRES) via the Trilinos subpackages Belos/Tpetra \cite{trilinos-website}. Despite the scale of these systems, GMRES consistently  converged in fewer than 20 iterations across all simulations presented herein, with the computational cost of each iteration dominated by a single FMM evaluation. In this work, we discretize the fiber centerlines on a Chebyshev grid with four quadrature points. While this is a coarse discretization, it reflects the assumption that long-range hydrodynamic effects will dominate near-field hydrodynamics.

\section{Correlation functions} \label{sec:corr}

Ideally, we want to compute quantities from the particle configuration without reference to an underlying grid. The reason is two-fold: (1) this is more efficient than evaluating the velocity field at a set of grid points, which additionally requires removal of singularities caused by grid points that inevitably lie inside of a particle, and (2) it limits bias from choosing coarse-graining boxes.

\subsection{Configuration variables}

Recall the definition of the empirical distribution
\begin{equation}
    \Psi_N(\bx, \bp, t) = \frac{|V|}{N} \sum_{n = 1}^N \delta_{\bx_n, \bp_n}
\end{equation}
and its associated order parameters $c = [1]_S$, $\bn = [\bp]_S$, and $\bQ = [\bp \bp - \bI / 3 ]_S$, where $[f]_S = \int_S f \Psi ~ dS_\bp$. Each of these fields takes the form
\begin{equation}
    \Phi(\bx) = \frac{|V|}{N}\sum_{n=1}^N \phi(\bp_n)\delta_{\bx_n},
\end{equation}
whose Fourier transform is
\begin{equation}
    \tilde\Phi(\bk) = \frac{1}{N}\sum_{n=1}^N \phi(\bp_n)e^{-i\bk\cdot\bx_n}.
\end{equation}
This sum can be computed efficiently using the Non-Uniform Fast Fast Fourier Transform (NUFFT) \cite{Barnett2019, Barnett2021}. In the NUFFT, we use $K = \lfloor (L/\ell) / 2 \rfloor$ target modes in each spatial dimension, which scales with the relative domain size so that the smallest length scale resolved is consistent across simulations. We find amplitudes of the correlation functions are somewhat sensitive to $K$, however the overall behavior is preserved.

The NUFFT can also be used to efficiently compute the radially averaged autocorrelation function, which in real space is defined as
\begin{equation}
    \Corr[\Phi](r) = \frac{1}{|S|} \int_S \left(\frac{1}{|V|}\int_V \Phi(\bx + r\hat\br)\Phi(\bx) ~ dV_\bx \right) ~ dS_{\hat\br}.
\end{equation}
For vector or tensor valued functions, the correlation function is summed over all elements, that is $\Corr[\bn] = \sum_{i = 1}^3 \Corr[n_i]$ and $\Corr[\bQ] = \sum_{i,j = 1}^3 \Corr[Q_{ij}]$. Given the coefficients $\tilde\Phi$ from the NUFFT, the correlation function can be defined by
\begin{equation}
    \Corr[\Phi](r) = \frac{1}{|S|} \int_S \cF^{-1}\big[|\tilde\Phi|^2 \big ](r \hat\br) ~ dS_{\hat\br}.
\end{equation}
We similarly identify the $L^2(V)$ norm with $||\Phi||_2^2 = \lim_{r\rightarrow0}\Corr(r)$. In practice the values of $\cF^{-1}\big[|\tilde\Phi|^2 \big ]$ are known on a Cartesian grid. To compute the radial average, we bin grid points into annuli of width $(L/\ell) / K$ and average over each annulus.

\subsection{Velocity}

Here we show an efficient method for computing the velocity spectrum from the empirical distribution. As described above, neglecting the lower-order doublet contribution in the RPY kernel, the velocity field generated by the collection of particles is
\begin{equation}
    \bu(\bx) = \frac{1}{8\pi\mu}\sum_{n = 1}^N [ \bG(\bx - (\bx_n + s\bp_n))\cdot\bff_n(s)]_\ell,
\end{equation}
and its Fourier transform is
\begin{widetext}
\begin{equation}
    \tilde \bu(\bk) = \frac{1}{8\pi\mu |V|} \left(\int_V \sum_{n = 1}^N \Big[  e^{-i\bk\cdot\bx} \bG(\bx - (\bx_n + s\bp_n)) \cdot\bff_n(s) \Big]_\ell ~ dV_\bx\right)
\end{equation}
\end{widetext}
Making the change of variables $\bx' = \bx - (\bx_n + s\bp_n)$ gives
\begin{equation}
    \tilde \bu(\bk) = \frac{\tilde \bG(\bk)}{8\pi\mu |V|}\cdot\left(\sum_{n = 1}^N \left[ e^{-i\bk \cdot (\bx_n + s\bp_n)} \bff_n(s) \right]_\ell\right),
\end{equation}
where $\tilde \bG(\bk) = 8\pi(\bI - \hat\bk\hat\bk)/k^2$ is the Fourier transform of the Stokeslet. In the numerical implementation, the centerline integral is discretized over quadrature points $s_m$ with weights $w_m$, giving
\begin{equation}
    \tilde \bu(\bk) = \frac{\tilde \bG(\bk)}{8\pi\mu |V|}\cdot\left(\sum_{n = 1}^N \sum_{m = 1}^M e^{-i\bk \cdot \by_{n,m}} \bc_{n, m} \right),
\end{equation}
which can be computed efficiently using a type 1 NUFFT with strengths $\bc_{n,m} = \bff_n(s_m)w_m$ at points $\by_{n,m} = \bx_n + s_m\bp_n$. From this spectrum we define the correlation function and $L^2$ norm as before. 

\section{Mean-field limit}\label{sec:mean_field}

In this section we formally derive the mean-field model in the limit $\ell / L \rightarrow 0$; similar derivations can be found in various references \cite{SS2013,Duerinckx2024}. All variables are in dimensional form to highlight their physical values, and we assume $L$ is fixed and treat $\ell$ as a small parameter.

\subsection{Stress tensor}

The stress in the suspension can be computed through Batchelor's formula \cite{Batchelor1970, GuazzelliMorris2011}, which relates the surface forces on a collection of particles within a control volume $B_\veps(\bx)$ centered at $\bx$ to a macroscopic averaged stress tensor,
\begin{equation}
    \obSigma(\bx) = -\frac{1}{|B_\veps(\bx)|}\sum_{\bx_n \in B_\veps(\bx)} \int_{S_n} (\bsigma_n \cdot \hat \bn)\by ~ dS_\by,
\end{equation}
where $S_n$ is the surface of particle $n$, $\bsigma_n$ is the stress tensor, and $\hat\bn$ is the normal vector. In terms of the singularity distribution, this is given by
\begin{equation}
    \obSigma(\bx) =
    -\frac{1}{|B_\veps(\bx)|} 
    \sum_{\bx_n \in B_\veps(\bx)} [ s\bff_n \bp_n ]_\ell.
\end{equation}
Multiplying the force balance Eq. (\ref{eq:sbt_balance}) by $\bI - \bp_n\bp_n/2$ gives the force distribution,
\begin{equation}
    \eta \bff_n(s) = (\bI - \bp_n\bp_n/2)\cdot(\dot\bx_n + s\dot\bp_n + u_n^s\bp_n - \bu_n),
\end{equation}
where as above we use the shorthand notation $\bu_n(s) = \bu(\bx_n + s\bp_n)$. Because the $\dot\bx_n$ term does not depend on $s$, it will vanish when $s\dot\bx_n$ is integrated over the particle centerline. We compute the remaining terms separately. Using the solution for $\dot\bp_n$ in Eq. (\ref{eq:pdot_n}), the rotational term $s\dot\bp_n\bp_n$ integrates to $(\bI - \bp_n\bp_n)\cdot[ s\bu\bp_n]_\ell$. Similarly, the slip term is $(U\ell^2/8)\bp_n\bp_n$ and the advective term $(\bI - \bp_n\bp_n/2)\cdot[ s\bu_n\bp_n]_\ell$. Taylor expanding $\bu_n(s) = \bu\rvert_{\bx_n} + s\bp_n\cdot\grad \bu\rvert_{\bx_n} + O(\ell^2)$, assuming the particles are well separated such that this series converges, and taking $|B_\veps|\rightarrow 0$ yields the mean-field stress tensor
\begin{equation}
\obSigma = \sigma_a\obQ + \sigma_b \obS : \obE,
\end{equation}
where $\sigma_a = -\beta\nu/8\eta$ and $\sigma_b = \nu/24\eta$ are constants. This stress balances the averaged viscous stress, treated through a similar procedure, yielding the Stokes equations for the mean-field velocity and pressure
\begin{gather}
    -\mu\Delta\obu + \grad \overline q = \div\obSigma,\\
    \div\obu = 0.
\end{gather}
Note that without the scaling $U \sim \ell$ the dipole coefficient $\sigma_a$ will not converge to a finite value.

\subsection{Smoluchowski equation}

Linearizing the velocity about the particle centerline $\bu_n(s) = \bu\rvert_{\bx_n} + {s\bp_n\cdot\grad_{\bx}\bu\rvert_{\bx_n}} + O(\ell^2)$ as above, again assuming particles are well-separated, the $O(\ell^2)$ center of mass dynamics are
\begin{align}
    \dot \bx_n &= \bu\rvert_{\bx_n} + \ell\beta\bp_n + O(\ell^2),\label{eq:xdot_n_linear}\\
    \dot \bp_n &= (\bI - \bp_n\bp_n)\cdot(\bp_n\cdot \grad\bu\rvert_{\bx_n}) + O(\ell^2).\label{eq:pdot_n_linear}
\end{align}
For any $N$, the empirical distribution $\Psi_N$ satisfies, in a weak sense, the Smoluchowski equation subject to these fluxes,
\begin{equation}
    \frac{\partial \Psi_N}{\partial t} + \grad_\bx \cdot(\dot\bx_N \Psi_N) + \grad_\bp \cdot(\dot \bp_N \Psi_N) = 0, 
\end{equation}
where $\dot\bx_N(\bx, \bp, t)$ and $\dot\bp_N(\bx, \bp, t)$ are functions that agree with $\dot\bx_n$ and $\dot\bp_n$ at the points $(\bx_n, \bp_n)$ for $n = 1, \ldots, N$. Taking $\ell\rightarrow 0$, and correspondingly $\Psi_N\rightarrow \Psi$, we get
\begin{equation}
    \dot\bx = \overline{\bu},\quad
    \dot \bp = (\bI - \bp\bp)\cdot(\bp\cdot \grad\overline{\bu}),\label{eq:pdot_linear}
\end{equation}
along with the Smoluchowski equation
\begin{equation}
    \frac{\partial\Psi}{\partial t} + \grad_\bx\cdot(\dot\bx\Psi) + \grad_\bp\cdot(\dot\bp\Psi) = 0.
\end{equation}

\subsection{Numerical implementation}

We discretize the mean-field PDEs using a pseudo-spectral method with $256^3$ Fourier modes, along with a second-order implicit-explicit multistep method for time-stepping. The time step size is chosen adaptively with respect to the maximal flow speed $\norm{\obu}_\infty$ and a CFL condition, typically taking values of $O(10^{-2})$. The contraction $\obS_B : \obE$ is computed using the fast method described in Ref. \cite{Weady2022} with Chebyshev interpolants of degree 80. In practice, a diffusive term of the form $D \Delta \obQ$ with $D = 10^{-4}$ is added to ensure the numerical solution is sufficiently resolved. The variable coefficient constraint stress in the Stokes equation is determined through fixed point iteration, which we find converges to $O(10^{-8})$ in roughly 5 iterations for our choice of $\zeta = 0.1421$. 

\bibliography{refs}

\end{document}